\documentclass[twoside,fleqn]{article}
\usepackage{espcrc2,epsfig}
\newcommand{\cswv}{{\rm c_{sw}}}
\newcommand{\dst}{\displaystyle}
\newcommand{\csw}{{\rm c_{swr}}}
\newcommand{\AmS}{{\protect\the\textfont2
  A\kern-.1667em\lower.5ex\hbox{M}\kern-.125emS}}
\title{
\rightline{HUB-EP-96/39}\vspace{1cm}
    Perturbative renormalisation of bilinear quark and
    gluon operators\thanks{Contribution presented by H. Perlt}
}
\author{
    M.~G\"ockeler$^{\rm a,b,c}$, 
    R.~Horsley$^{\rm d}$, 
    E.-M.~Ilgenfritz$^{\rm d}$, 
    H.~Oelrich$^{\rm a,e}$,
    H.~Perlt$^{\rm f}$,
    P.~Rakow$^{\rm a,e}$,
    G.~Schierholz$^{\rm a,g}$,
    A.~Schiller$^{\rm f}$ and 
    P.~Stephenson$^{\rm e}$ \\[1em] 
{$^{\rm a}$ HLRZ, c/o Forschungszentrum J\"ulich, D-52425 J\"ulich, Germany} \\[0.5em]
{$^{\rm b}$ Institut f\"ur Theoretische Physik, RWTH Aachen, D-52056 Aachen, Germany} \\[0.5em]
{$^{\rm c}$ Institut f\"ur Theoretische Physik, J.~W.~Goethe
Universit\"at, D-60054 Frankfurt, Germany} \\[0.5em]
{$^{\rm d}$ Institut f\"ur Physik, Humboldt-Universit\"at zu Berlin, D-10115 Berlin,
Germany}\\[0.5em]
{$^{\rm e}$ DESY-IfH Zeuthen, D-15735 Zeuthen, Germany} \\[0.5em]
{$^{\rm f}$ Institut f\"ur Theoretische Physik, Universit\"at Leipzig,
D-04109 Leipzig, Germany}\\[0.5em]
{$^{\rm g}$ DESY, D-22603 Hamburg, Germany}\\
}
\begin{document}
\begin{abstract}
 The renormalisation constants for local bilinear quark operators are
calculated using the Sheikholeslami-Wohlert improved action. In addition
we compute the renormalisation constant of the leading gluon operator
for different group representations and discuss the mixing of the operators
$\vec{E}^2$ and $\vec{B}^2$.
\end{abstract}

\maketitle
\section{INTRODUCTION}

Forward hadron matrix elements of bilinear quark and gluon
operators, as they appear e.g. in the operator product expansion for the nucleon
structure functions, have been calculated with some success on the
lattice \cite{meinulf96}. In order to extract the relevant physical
information from the lattice matrix elements one has to renormalise the
operators. This can be done perturbatively \cite{gqcd2,rossi,capi,poch} or
non-perturbatively \cite{martinelli,gqcd1}. Both methods
have their advantages and disadvantages. In this paper we present a
one-loop calculation of the renormalisation constants of some bilinear quark
operators for improved Wilson fermions and of the gluon operator
${\rm Tr} F_{\mu\rho}F_{\rho\nu}$. All
calculations have been performed in the quenched approximation and for
zero renormalised quark mass.

\section{LOCAL QUARK OPERATORS}

Lattice calculations with Wilson fermions suffer from cut-off effects of
${\cal O}(a)$, where $a$ is the lattice
spacing. By adding the local counterterm
\begin{equation}
\Delta S = -i \cswv \; g a^4 \sum_{n,\mu\nu} \frac{a r}{4} \overline{\psi}_n
 \sigma_{\mu\nu} F_{n,\mu\nu} \psi_n
 \label{SW}
\end{equation}
to the action, cut-off effects in on-shell quantities can be reduced to
${\cal O}(a^2)$ \cite{sw} for a particular choice of $\cswv$.

In hadron matrix element calculations one has to improve the operators as
well. This again can be achieved by adding certain counterterms. For a
class of operators, in particular for the operators considered in this paper,
the counterterms can be obtained by a rotation of the fermion fields
\[
\overline{\psi} \rightarrow \overline{\psi}(1+\csw\;\frac{ar}{4}
\stackrel{\leftarrow}{D}), \quad
\psi \rightarrow (1- \csw\;\frac{ar}{4}\stackrel{\rightarrow}{D})\psi,
\]
with $\csw$ to be determined properly.

In lowest order perturbation theory $\cswv = \csw = 1$. A non-perturbative
evaluation of $\cswv$ was given in \cite{Sommer}. Throughout the rest of the
paper we shall assume $r = 1$

We consider the operators
\[
{\cal O} = \bar{\psi} \Gamma \psi
\]
with
\[
\Gamma = 1,\; \gamma_5,\; \gamma_\mu,\; \gamma_\mu\gamma_5,\;
\sigma_{\mu\nu}\gamma_5.
\]
The renormalisation constants are generically defined by (for details see
\cite{gqcd2})
\[
{\cal O}(\mu) = Z_{\cal O}(a\mu,g) {\cal O}(a),
\]
\begin{equation}
\langle q(p)|{\cal O}(\mu)|q(p)\rangle =
\langle q(p)|{\cal O}(a)|q(p)\rangle \big|^{tree}_{p^2=\mu^2}.
\label{eq1}
\end{equation}
They can be cast into the form
\[
Z_{\cal O}=1 - \frac{\displaystyle g^2}{\displaystyle 16\pi^2}
C_F\big(\gamma_{\cal O} \ln(a\mu) + B_{\cal O}\big), \;
C_F=\frac{4}{3},
\]
where $\gamma_{\cal O}$ is the anomalous dimension and $B_{\cal O}$ is the
finite part of the renormalisation constant. The definition (\ref{eq1})
corresponds to the momentum subtraction scheme.

$B_{\cal O}$ receives contributions from two different sources: the proper
operator radiative corrections and the self-energy diagrams. We write
accordingly
\[
B_{\cal O} = B^{proper}_{\cal O} + B^{self}_{\cal O}.
\]
Using the procedure outlined in \cite{gqcd2} we obtain~\footnote{The numbers
are given here only to three decimal places but are known to more than
ten places.}
\begin{eqnarray*}
B^{proper}_1 & = & 6.100 - 19.172\;\csw - 4.981\;\csw^2 + \\
            &   & 9.987\;\cswv + 13.801\;\csw\;\cswv + \\
            &   & 2.177\;\csw^2\;\cswv + 0.017\;\cswv^2 - \\
            &   & 3.538\;\csw\;\cswv^2 - 0.288\;\csw^2\;\cswv^2 + \\
            &   & (-2 + \gamma_E - F_0) \;\xi ,\\
B^{proper}_{\gamma_5}  & = &  15.743 + 8.660\;\csw^2 - \\
              &   &  5.715\;\csw^2\;\cswv + 3.433\;\cswv^2 + \\
              &   &  1.361\;\csw^2\;\cswv^2 + \\
              &   & (-2 + \gamma_E - F_0)\;\xi ,\\
B^{proper}_{\gamma_\mu}  & = &  8.765 - 9.786\;\csw + 3.525\;\csw^2 - \\
                &   &  2.497\;\cswv + 3.416\;\csw\;\cswv - \\
                &   &  1.973\;\csw^2\;\cswv + 0.854\;\cswv^2 + \\
                &   &  0.885\;\csw\;\cswv^2 +  0.412\;\csw^2\;\cswv^2 + \\
                &   &  (-1 + \gamma_E - F_0)\;\xi , \\
B^{proper}_{\gamma_\mu\gamma_5}  & = &  3.944 - 19.372\;\csw - 3.296\;\csw^2 + \\
                        &   &  2.497\;\cswv + 10.317\;\csw\;\cswv + \\
                        &   &  1.973\;\csw^2\;\cswv - 0.854\;\cswv^2 - \\
                        &   &  0.885\;\csw\;\cswv^2 - 0.412\;\csw^2\;\cswv^2 + \\
                        &   &  (-1 + \gamma_E - F_0)\;\xi , \\
B^{proper}_{\sigma_{\mu \nu} \gamma_5}  & = & 4.166 - 16.243\;\csw
- 0.461\;\csw^2 - \\
         &   & 1.664\;\cswv +  6.855\;\csw\;\cswv + \\
         &   & 0.590\;\csw^2\;\cswv - 0.575\;\cswv^2 + \\
         &   & 0.590\;\csw\;\cswv^2 - 0.179\;\csw^2\;\cswv^2 + \\
         &   & (\gamma_E - F_0)\;\xi.
\end{eqnarray*}
Here
$\xi$ is the gauge parameter ($\xi=0$ for Feynman gauge, $\xi=1$ for Landau
gauge) and $F_0 = 4.369225$.
The self-energy contribution is
\begin{eqnarray*}
B^{self}_{\cal O} & = & -0.381 - 2.249\;\cswv - 1.397\;\cswv^2 + \\
              &   & 8\pi^2\;Z_0 + (1-\gamma_E + F_0)\;\xi,
\end{eqnarray*}
where $Z_0 = 0.154933$.

Since the Wilson coefficients are
usually computed in the MS or $\overline{{\rm MS}}$ scheme,
one would like to know the renormalisation constants in these schemes too.
The transformations between the different schemes are \cite{gqcd2}
\[
B_{\cal O}^{MS} = B_{\cal O} - B_{\cal O}^{con},
\]
\[
B_{\cal O}^{\overline{MS}} = B_{\cal O}^{MS}   +
\frac{\gamma_{\cal O}}{2}(\gamma_E - \ln 4\pi),
\]
where $\gamma_{\cal O}$ and $B_{\cal O}^{con}$, the finite contribution of
the continuum integrals to $B_{\cal O}$, are given in
the table below:

\begin{table}[h]
\label{anomdimtab}
\vspace{0.5cm}
\begin{tabular}{||c|c|c||}
\hline
 ${\cal O}$&   $\gamma_{\cal O}$ &  $B_{\cal O}^{con}$   \\
\hline
$1$, $\gamma_5$   & $-6 $ & $5 +
\frac{\dst\gamma_{\cal O}}{\dst 2} c_E - \xi$ \\
[0.7ex]
  $\gamma_\mu$, $\gamma_\mu\gamma_5$    & $0$ & $0$\\
[0.7ex]
$\sigma_{\mu \nu} \gamma_5 $        & $ 2 $ & $-1 + 
\frac{\dst\gamma_{\cal O}}{\dst 2}c_E + \xi$   \\
[0.7ex]
\hline
\end{tabular}
\end{table}

\noindent
Here $c_E = \gamma_E - \ln 4\pi$.
In the MS and in the $\overline{{\rm MS}}$ scheme $B_{\cal O}$ is
gauge independent, as one can easily see.

Our results for $\cswv = 1, \csw = 0$ and $\cswv = \csw = 1$ agree with those
of \cite{burgio} and \cite{borelli}, respectively.

\section{GLUON OPERATOR}

Several information about the gluon content of the nucleon can be derived
from the matrix element of the operator \cite{roger96}
\begin{equation}
{\cal O}_{\mu\nu} = {\rm Tr} F_{\mu\rho}F_{\rho\nu} .
\label{gluon1}
\end{equation}
The renormalisation constants of certain irreducible representations of
(\ref{gluon1}) have been given elsewhere \cite{menotti,capitani,gqcd3}. Here
we are interested in the operators
\[
{\rm Tr}\vec{E}^2 = - {\cal O}_{44},\;  {\rm Tr}\vec{B}^2 =
\frac{1}{2}({\cal O}_{44} - \sum_{i=1}^3{\cal O}_{ii}),
\]
where
\[
{\rm E}_i = {\rm F}_{i4}, \;
{\rm B}_i = \frac{1}{2} \epsilon_{ijk}{\rm F}_{jk}.
\]
It is clear from group theoretical arguments \cite{gqcd4} that
${\rm Tr}\vec{E}^2$ and ${\rm Tr}\vec{B}^2$ mix under renormalisation.
We write in symbolic form
\begin{eqnarray*}
{\rm Tr}\vec{E}^2_R & = & Z_{EE}\; {\rm Tr}\vec{E}^2_0 + Z_{EB} \; {\rm Tr}\vec{B}^2_0 \\
{\rm Tr}\vec{B}^2_R & = & Z_{BE}\; {\rm Tr}\vec{E}^2_0 + Z_{BB} \; {\rm Tr}\vec{B}^2_0.
\end{eqnarray*}
In the quenched approximation all $Z$'s are finite, i.e. have no
logarithmic contributions. Symmetry arguments
lead to the following relations:
\[
Z_{EE} = Z_{BB}, \; Z_{EB} = Z_{BE}.
\]
Making use of some of the results of \cite{menotti} we obtain in
the MS scheme
\begin{eqnarray*}
Z_{EE}  =  Z_{BB} &=& 1 +  0.123128 g^2 \\
Z_{EB}  =  Z_{BE} &=& -0.096655 g^2.
\end{eqnarray*}

There is a combination of ${\rm Tr}\vec{E}^2$ and ${\rm Tr}\vec{B}^2$
which belongs to an irreducible representation of the hypercubic group and
does not mix:
\begin{eqnarray*}
{\cal O}_b&=&{\cal O}_{44} - \frac{1}{3}({\cal O}_{11}+{\cal O}_{22}+
{\cal O}_{33})\\
& =& \frac{2}{3}({\rm Tr}\vec{B}^2 -{\rm Tr}\vec{E}^2) .
\end{eqnarray*}
For the renormalisation constant of this operator we find
\[
Z_{{\cal O}_b} = 1 +  0.219783 g^2
\]
which confirms numerically the relation between the $Z$'s,
\[
Z_{{\cal O}_b} = Z_{EE} - Z_{EB}.
\]
The renormalization
constant  $Z_{{\cal O}_b}$ should be compared with the renormalisation constant
of the operator
\begin{eqnarray*}
{\cal O}_a = {\cal O}_{\{i4\}} =  {\rm Tr} (\vec{E} \times \vec{B})_i
\end{eqnarray*}
which turns out to be \cite{capitani,gqcd3}
\[
Z_{{\cal O}_a} = 1 + 0.27075 g^2.
\]
(The lattice value is $1+0.29608 g^2$ for $\xi = 0$.) The mall difference of
$Z_{{\cal O}_a}$ and $Z_{{\cal O}_b}$ is due to
non-$O(4)$ invariant contributions.

\end{document}